\documentclass[a4paper,fleqn,usenatbib]{mnras}

\usepackage[T1]{fontenc}
\usepackage{ae,aecompl}
\usepackage{graphicx}	
\usepackage{amsmath}	
\usepackage{amssymb}	

\newcommand{\km}{\mathrm{km}}
\newcommand{\pc}{\mathrm{pc}}
\newcommand{\AU}{\mbox{AU}}
\newcommand{\s}{\mathrm{s}}

\newcommand{\Gauss}{\mathrm{G}}
\newcommand{\cm}{\mbox{cm}}
\newcommand{\g}{\mbox{g}}

\newcommand{\mach}{\mathcal{M}_\mathrm{S}}
\newcommand{\macha}{\mathcal{M}_\mathrm{A}}

\newcommand{\cs}{c_\mathrm{s}}
\newcommand{\vect}[1]{{\textbf{\textit{#1}}}}

\title[Synchrotron probes of turbulent driving.]{Probes of turbulent driving mechanisms in molecular clouds from fluctuations in synchrotron intensity}

\author[Herron et al.]{
C.~A.~Herron$^{1}$\thanks{E-mail: C.Herron@physics.usyd.edu.au (CAH)},
C.~Federrath$^{2}$,
B.~M.~Gaensler$^{3,1}$,
G.~F.~Lewis$^{1}$,
\newauthor{N.~M.~McClure-Griffiths$^{2}$, Blakesley Burkhart$^{4}$}
\\
$^{1}$Sydney Institute for Astronomy, School of Physics, A28, The University of Sydney, NSW, 2006, Australia\\
$^{2}$Research School of Astronomy and Astrophysics, Australian National University, Canberra, ACT 2611, Australia\\
$^{3}$Dunlap Institute for Astronomy and Astrophysics, University of Toronto, 50 St. George Street, Toronto, Ontario, M5S 3H4, Canada\\
$^{4}$Harvard-Smithsonian Center for Astrophysics, 60 Garden St. Cambridge, MA, USA
}

\date{Accepted XXX. Received YYY; in original form ZZZ}

\pubyear{2016}

\begin{document}
\label{firstpage}
\pagerange{\pageref{firstpage}--\pageref{lastpage}}
\maketitle

\begin{abstract}
Previous studies have shown that star formation depends on the driving of molecular cloud turbulence, and differences in the driving can produce an order of magnitude difference in the star formation rate. The turbulent driving is characterised by the parameter $\zeta$, with $\zeta=0$ for compressive, curl-free driving (e.g. accretion or supernova explosions), and $\zeta=1$ for solenoidal, divergence-free driving (e.g. Galactic shear). Here we develop a new method to measure $\zeta$ from observations of synchrotron emission from molecular clouds. We calculate statistics of mock synchrotron intensity images produced from magnetohydrodynamic simulations of molecular clouds, in which the driving was controlled to produce different values of $\zeta$. We find that the mean and standard deviation of the log-normalised synchrotron intensity are sensitive to $\zeta$, for values of $\zeta$ between $0$ (curl-free driving) and $0.5$ (naturally-mixed driving). We quantify the dependence of zeta on the direction of the magnetic field relative to the line of sight. We provide best-fit formulae for $\zeta$ in terms of the log-normalised mean and standard deviation of synchrotron intensity, with which $\zeta$ can be determined for molecular clouds that have similar Alfv\'enic Mach number to our simulations. These formulae are independent of the sonic Mach number. Signal-to-noise ratios larger than $5$, and angular resolutions smaller than $5\%$ of the cloud diameter, are required to apply these formulae. Although there are no firm detections of synchrotron emission from molecular clouds, by combining Green Bank Telescope and Very Large Array observations it should be possible to detect synchrotron emission from molecular clouds, thereby constraining the value of $\zeta$.

\end{abstract}




\section{Introduction} \label{Intro}
Turbulence is ubiquitous throughout the Milky Way, and plays a pivotal role in galactic processes such as star formation \citep{Ferriere2001, Elmegreen2004, McKee2007, Padoan2014}. There are numerous mechanisms believed to be responsible for driving interstellar turbulence (see \citealt{Elmegreen2004}, \citealt{MacLow2004}, and \citealt{Federrath2016} for reviews), including galaxy mergers \citep{Renaud2014}, the gravitational potential of spiral arms \citep{Falceta2015}, differential galactic rotation and baroclinicity (misaligned pressure and density gradients, \cite{DelSordo2011}), supernova explosions \citep{Hennebelle2014, Padoan2016}, stellar feedback \citep{Lee2012}, gravitational collapse and accretion \citep{Klessen2010, Krumholz2010, Federrath2011, Robertson2012, Krumholz2016}, and magnetohydrodynamic instabilities \citep{Tamburro2009}. 

Not only do these different driving mechanisms act on different spatial scales, but they also differ in how they drive turbulence. Some mechanisms, such as supernovae and gravitational collapse, tend to produce shocks and rarefactions, injecting energy into the compressible modes of the turbulence. For compressible modes, the curl of the driving force is zero. Conversely, shear due to the Galactic differential rotation creates vortices, and this represents an injection of energy into the solenoidal modes of the turbulence (\citealt{Krumholz2015} provide a possible example of this in the Central Molecular Zone, also see \citealt{Federrath2016b}). For solenoidal modes, the divergence of the driving force is zero. Whilst compressive and solenoidal driving forces have distinct properties, the velocity fields they produce are less distinct. In particular, the exponents of the velocity power spectra are similar for compressive and solenoidal driving \citep{Federrath2013}, and compressive driving can produce a velocity field that has energy in both compressive and solenoidal motions, and vice versa (see \citealt{Federrath2010}, Figure 14, \citealt{Kowal2010}, Figure 2). For example, supernovae primarily act as compressive drivers of turbulence, while viscosity, baroclinicity, shear, and rotation generate vorticity in the motion of the gas, transferring energy into solenoidal fluid motions \citep{Federrath2011b, DelSordo2011, Padoan2016}. The energy in solenoidal motions was quantified by \cite{Federrath2011b} (bottom of their Figure 3), for simulations with compressive and solenoidal driving, and a range of sonic Mach numbers. They found that for supersonic simulations, solenoidal and compressive driving would cause approximately $80\%$ and $40\%$ of the kinetic energy to be in solenoidal motions, respectively, due to the transferral of energy between solenoidal and compressive motions.

Whether turbulence is driven compressively or solenoidally has significant implications for how gas in molecular clouds evolves. For instance, \cite{Federrath2012} found that the star formation rate in their simulations was ten times larger for simulations with compressive driving compared to simulations with solenoidal driving of the turbulence. Furthermore, \cite{Renaud2014} found in their simulations that galaxy mergers increased the power in compressive fluid motions, leading to enhanced star formation. As compressive driving is associated with gravitational collapse of gas and supernovae, it can also be associated with star formation that has occurred recently or will occur soon. Solenoidal driving, on the other hand, tends to be associated with quiescent gas \citep{Federrath2010}. Hence, the relative strength of the compressive and solenoidal driving determines the structure and star formation in molecular clouds.

To describe the relative strengths of the compressive and solenoidal modes, \cite{Federrath2010} introduced the turbulent driving parameter $\zeta$. From this parameter, the ratio of power in compressive forcing modes, $F_{\mathrm{comp}}$, relative to the total forcing power, $F_{\mathrm{total}} = F_{\mathrm{comp}} + F_{\mathrm{sol}}$, where $F_{\mathrm{sol}}$ is the power in solenoidal forcing modes, can be determined:
\begin{equation}
\frac{F_{\mathrm{comp}}}{F_{\mathrm{total}}} = \frac{(1-\zeta)^2}{1-2\zeta+D\zeta^2} \label{zeta_def}
\end{equation}
Equation \ref{zeta_def} defines $\zeta$, where $D$ is the number of spatial dimensions (see Figure 1 of \citealt{Federrath2010}). Here we only consider $D=3$. $\zeta$ is defined such that $\zeta \in [0,1]$, where $\zeta=0$ for purely compressive driving, and $\zeta=1$ for purely solenoidal driving. `Natural mixing' of the turbulence corresponds to one-third of the injected energy going into compressive modes, as these modes are longitudinal and occupy one of the three spatial dimensions, and in this case $\zeta = 0.5$. 

In previous observational \citep{RomanDuval2011, Ginsburg2013, Kainulainen2013} and theoretical \citep{Federrath2008, Federrath2009, Price2011, Burkhart2012, Konstandin2012, KonstandinEtAl2012, Micic2012, Federrath2013} studies, it has been found that compressive driving produces larger density enhancements, a wider density probability distribution function (PDF), and larger voids than solenoidal driving, whereas solenoidal driving produces more vorticity. In particular, \cite{Federrath2010} showed that the standard deviation of the column density PDF was three times larger for compressive driving than for solenoidal driving, providing an observational method to probe $\zeta$. \cite{Brunt2014} also described a method to measure the momentum in solenoidal modes, which involves the power spectrum of spectral line intensities, but assumes statistical isotropy.

In this paper, we investigate whether statistics of synchrotron intensity can be used to constrain the turbulent driving parameter $\zeta$. Synchrotron emission is emitted by ultra-relativistic electrons as they spiral around magnetic field lines, and the intensity $I$ observed at frequency $\nu$ is given by \citep{Ginzburg1965}
\begin{align}
I(\nu) &= \frac{e^3}{4 \pi m_e c^2} \int_0^{L} \frac{\sqrt{3}}{2 - 2 \alpha} \Gamma \left( \frac{2 - 6 \alpha}{12} \right) \Gamma \left( \frac{22 - 6 \alpha}{12} \right) \times \nonumber \\ 
& \quad \quad \left( \frac{3e}{2 \pi m_e^3 c^5} \right)^{-\alpha} K B_{\perp}^{1-\alpha} \nu^{\alpha} \, \mathrm{d} L', \label{sync_inten}
\end{align}
for a path length $L$ through a synchrotron emitting medium. In Equation \ref{sync_inten}, $e$ is the charge of an electron, $m_e$ is the mass of an electron, $c$ is the speed of light, $\alpha$ is the spectral index of the emission, $K$ is a constant, $B_{\perp}$ is the magnetic field strength perpendicular to the line of sight, and $\Gamma$ denotes the gamma function. We do not specify a value for $K$, as it will cancel out when we calculate the log-normalised synchrotron intensity in Section \ref{sec_method}. In the derivation of Equation \ref{sync_inten}, it is assumed that the ultra-relativistic electrons have a homogeneous and isotropic distribution described by the power-law
\begin{equation}
N(E) \, \mathrm{d} E = K E^{2 \alpha - 1} \, \mathrm{d} E, \label{energy_dist}
\end{equation}
where $N(E)$ is the number density of electrons with energies between $E$ and $E + \mathrm{d}E$.

Recently, \cite{Herron2016} found that statistics of synchrotron intensity are sensitive to the Alfv\'enic Mach number of fully ionised turbulent gas. In this paper we apply a similar methodology to determine whether statistics of synchrotron intensity are also sensitive to the turbulent driving parameter $\zeta$ in molecular clouds. However, the brightness of synchrotron emission from molecular clouds is currently an open question, and there are no firm detections of synchrotron emission originating within molecular clouds. For example,  \cite{Protheroe2008} and \cite{Jones2011} estimate the synchrotron intensity of the molecular cloud Sagittarius B to be detectable in their observations, however do not detect any synchrotron emission from the molecular cloud at $1384$ or $2368$ MHz. Conversely, \cite{Yusef-Zadeh2013} claim a detection of synchrotron emission from a molecular cloud near the Galactic centre at $74$ MHz, but as mentioned by \cite{Dickinson2015}, the misalignment of the synchrotron emission and molecular cloud mean that the synchrotron emission cannot be confirmed to be arising from within the molecular cloud. Although there are currently no detections of synchrotron emission from molecular clouds, statistics of synchrotron intensity are promising because they may provide a method of determining $\zeta$ that is independent of the density in the emitting region, unlike the methods introduced by \cite{Ginsburg2013} and \cite{Brunt2014}.

In Section \ref{sec_sims} we describe our magnetohydrodynamic simulations, and in Section \ref{sec_method} we specify how we create mock synchrotron intensity images for our simulations, and what statistics of synchrotron intensity we calculate. Our results are presented in Section \ref{sec_results}, and discussed in Section \ref{sec_discuss}. In particular, we discuss the detectability of synchrotron emission from molecular clouds in Section \ref{observability}.

\section{Simulations} \label{sec_sims}
Our magnetohydrodynamical (MHD) simulations were performed with the adaptive mesh refinement \citep[AMR,][]{BergerColella1989}, multi-physics code FLASH \citep{FryxellEtAl2000,DubeyEtAl2008}, version~4. We use the robust positive-definite HLL3R Riemann scheme \citep{WaaganFederrathKlingenberg2011} to solve the compressible MHD equations on three-dimensional (3D) periodic grids of fixed side length $L$. The driving of the turbulence and the global simulation parameters are similar to the ones of \citet{Federrath2015}. Here we briefly summarise the turbulence driving module and the initial conditions.

\subsection{Turbulence driving} \label{sec:turbdriving}
We drive turbulence with the driving module developed by \citet{Federrath2010} and available in the public version of the FLASH code. The module drives turbulence similar to that observed in real molecular clouds, i.e., driving on large scales \citep{HeyerWilliamsBrunt2006,BruntHeyerMacLow2009}, producing a power-law velocity spectrum, $E(k)\sim k^{-2}$, consistent with the observed velocity dispersion--size relation \citep{Larson1981,HeyerBrunt2004,RomanDuval2011} and consistent with numerical simulations of supersonic turbulence \citep[][]{KritsukEtAl2007,SchmidtEtAl2009,Federrath2010,KonstandinEtAl2012,Federrath2013}. We use the stochastic Ornstein-Uhlenbeck process \citep{EswaranPope1988,SchmidtHillebrandtNiemeyer2006} to construct a random turbulent acceleration field ${\bf F}$ that varies smoothly in space and time on the autocorrelation timescale $t_\mathrm{turb}$. We set the autocorrelation timescale equal to the turbulent crossing time, $t_\mathrm{turb}=L/(2\sigma_v)$, where $\sigma_v$ is the velocity dispersion on large scales. Consistent with observational constraints, the driving only contains large-scale modes, $1<\left|\mathbf{k}\right|L/2\pi<3$, where most of the power is injected at the $k=2$ mode in Fourier space, which corresponds to half of the box size ($L/2$). The turbulence on smaller scales ($k\geq3$) is allowed to develop self-consistently through the non-linear interactions of the gas, governed by the MHD equations.

We decompose the driving field into a solenoidal and a compressive part by performing a Helmholtz decomposition with a projection operator in Fourier space. In index notation, the projection operator reads $\mathcal{P}_{ij}^\zeta\,(\vect{k}) = \zeta\,\mathcal{P}_{ij}^\perp+(1-\zeta)\,\mathcal{P}_{ij}^\parallel = \zeta\,\delta_{ij}+(1-2\zeta)\,k_i k_j/|\vect{k}|^2$, where $\mathcal{P}_{ij}^\perp$ and $\mathcal{P}_{ij}^\parallel$ are the solenoidal and compressive projection operators, respectively. This projection operation allows us to construct a solenoidal (divergence-free) or a compressive (curl-free) acceleration field by setting $\zeta=1$ or $\zeta=0$, respectively. We can further construct any arbitrary combination of solenoidal and compressive modes for the target driving field, by varying $\zeta$ in the range $[0,1]$ (see Equation \ref{zeta_def}). Here we compare five simulations with $\zeta=0$, $0.25$, $0.5$, $0.75$, and $1$, in order to study the dependence of the moments of the synchrotron radiation maps on the turbulence driving mode (see Table~\ref{tab:sims} for the list of simulation models).

\begin{table*}
\caption{Key simulation parameters and derived moments of the log-normalised synchrotron intensity.}
\label{tab:sims}
\def\arraystretch{1.1}
\setlength{\tabcolsep}{3.7pt}
\begin{tabular}{lcccccccc}
\hline
Model Name & $\zeta$ & $\mach$ & $\macha$ & LOS & $\mu_{\mathcal{I}}$ & $\sigma_{\mathcal{I}}$ & $\gamma_{\mathcal{I}}$ & $\beta_{\mathcal{I}}$ \\
(1) & (2) & (3) & (4) & (5) & (6) & (7) & (8) & (9) \\
\hline
Z1.00\_MS5\_MA2  & $1$ (sol) & $4.9\pm0.1$ & $2.1\pm0.1$ & x & -0.02 & 0.14 & -0.18 & 0.04\\
 &  &  &  & y & -0.03 & 0.15 & -0.22 & 0.24\\
 &  &  &  & z & -0.04 & 0.19 & -0.39 & 0.14\\
Z0.75\_MS5\_MA2  & $0.75$ & $4.9\pm0.2$ & $1.9\pm0.1$ & x & -0.03 & 0.15 & -0.12 & 0.01\\
 &  &  &  & y & -0.03 & 0.15 & -0.13 & -0.14\\
 &  &  &  & z & -0.03 & 0.17 & -0.12 & 0.15\\ 
Z0.50\_MS5\_MA2  & $0.5$ (mix) & $4.4\pm0.2$ & $2.0\pm0.2$ & x & -0.02 & 0.14 & 0.01 & -0.12\\
 &  &  &  & y & -0.02 & 0.14 & -0.26 & -0.12\\
 &  &  &  & z & -0.04 & 0.19 & -0.22 & 0.06\\
Z0.25\_MS5\_MA2  & $0.25$ & $4.7\pm0.2$ & $2.4\pm0.2$ & x & -0.05 & 0.20 & 0.11 & -0.23\\
 &  &  &  & y & -0.04 & 0.19 & -0.01 & -0.30\\
 &  &  &  & z & -0.11 & 0.32 & -0.26 & -0.09\\
Z0.00\_MS5\_MA2  & $0$ (comp) & $4.4\pm0.3$ & $2.3\pm0.4$ & x & -0.07 & 0.23 & 0.58 & 0.60\\
 &  &  &  & y & -0.07 & 0.24 & 0.46 & 0.50\\
 &  &  &  & z & -0.15 & 0.36 & 0.13 & -0.22\\
\hline
\end{tabular}
\begin{minipage}{\linewidth}
\textbf{Notes.} Column 1: simulation name. Column 2: type of turbulence driving, parameterised by $\zeta$ \citep{Federrath2010} (see Equation \ref{zeta_def}). Columns 3, 4: turbulent sonic and Alfv\'en Mach numbers; the 1-sigma uncertainties denote the variations of $\mach$ and $\macha$ in the regime of fully-developed turbulence. Column 5: line of sight used to produce the synchrotron intensity map. Columns 6, 7, 8, 9: mean, standard deviation, skewness, and kurtosis of the log-normalised synchrotron intensity.
\end{minipage}
\end{table*}

\subsection{Initial conditions and simulation parameters}
All our simulations start from the same uniform gas density, with uniform initial magnetic field in the $z$ direction of the computational domain and zero initial velocities. Each simulation has a box size of $L=2\,\pc$ and the gas has a mean density of $\rho_0=6.56\times10^{-21}\,\g\,\cm^{-3}$ and an initial magnetic field of $B=10\,\mu\Gauss$ along the z direction of the computational domain. The temperature is held fixed at $T=10\,\mathrm{K}$, leading to a constant sound speed of $\cs=0.2\,\km\,\s^{-1}$. After an initial transient phase that lasts for two turbulent crossing times, $t\leq2t_\mathrm{turb}$ \citep{Federrath2009,Federrath2010,Federrath2013}, during which the turbulence becomes fully developed, the RMS sonic and Alfv\'en Mach numbers are $\mach\sim5$ and $\macha\sim2$ and they fluctuate around these values at a $\sim10\%$ level. Table~\ref{tab:sims} summaries these key parameters for each simulation model. The simulation parameters were chosen to roughly follow typical Mach numbers measured in molecular clouds in the Milky Way \citep{FalgaronePugetPerault1992,FalgaroneEtAl2008,CrutcherEtAl2010,SchneiderEtAl2013}, and are approximately the same for all of our simulations, so that we only observe the effect of variations in the turbulent driving. If the Alfv\'enic Mach number of a turbulent region is significantly different to those we assume, then our results may not be applicable (see Appendix \ref{mach_dep}).

Each simulation uses a base grid of $256^3$ cells and we allow for 4 additional levels of AMR, giving a maximum effective resolution of $4096^3$ or a minimum cell size of $\Delta x = 100\,\AU$. The numerical mesh is adaptively refined based on the criterion that the Jeans length must be resolved with 16 to 32 grid cells at any point in space and time. This criterion guarantees that the turbulence and potential magnetic field amplification by dynamo action are sufficiently resolved \citep{Sur2010, Federrath2011}.


\section{Synthetic Observations of the Simulations} \label{sec_method}

To create synthetic images of the synchrotron intensity for the simulations, we follow the same procedure as \cite{Herron2016}, assuming a fixed observing frequency, and constant values of $\alpha$ and $K$ within the simulation cube. The consequence of these assumptions is that the observed synchrotron intensity is proportional to the integral of the magnetic field perpendicular to the line of sight, i.e.
\begin{equation}
I \propto \int B_{\perp}^{1-\alpha} \mathrm{d}L. \label{mock_sync}
\end{equation}

For each of the simulations, we produce maps of synchrotron intensity for lines of sight along the x, y, and z axes according to Equation \ref{mock_sync}, assuming $\alpha=-1$. For example, for a line of sight along the x axis, the magnetic field perpendicular to the line of sight is calculated as $B_{\perp} = \sqrt{B_y^2 + B_z^2}$. As the mean magnetic field is along the z axis, we expect the synchrotron intensity images to be different for a line of sight along this axis. To ensure that the intensity values obtained do not depend on the size of the simulation, we normalise the intensity at each pixel of the synchrotron map by the length of the simulation.

This process was repeated for five different temporal realisations of each simulation, in the regime of fully-developed turbulence, where the turbulence was allowed to evolve for half a turbulent crossing time between these realisations, i.e., at $t/t_{\mathrm{turb}}=2,2.5,3,3.5,$ and $4$. In total, there are $25$ mock images of synchrotron emission for each line of sight. 

\begin{figure*}
\begin{center}
\includegraphics[scale=0.75]{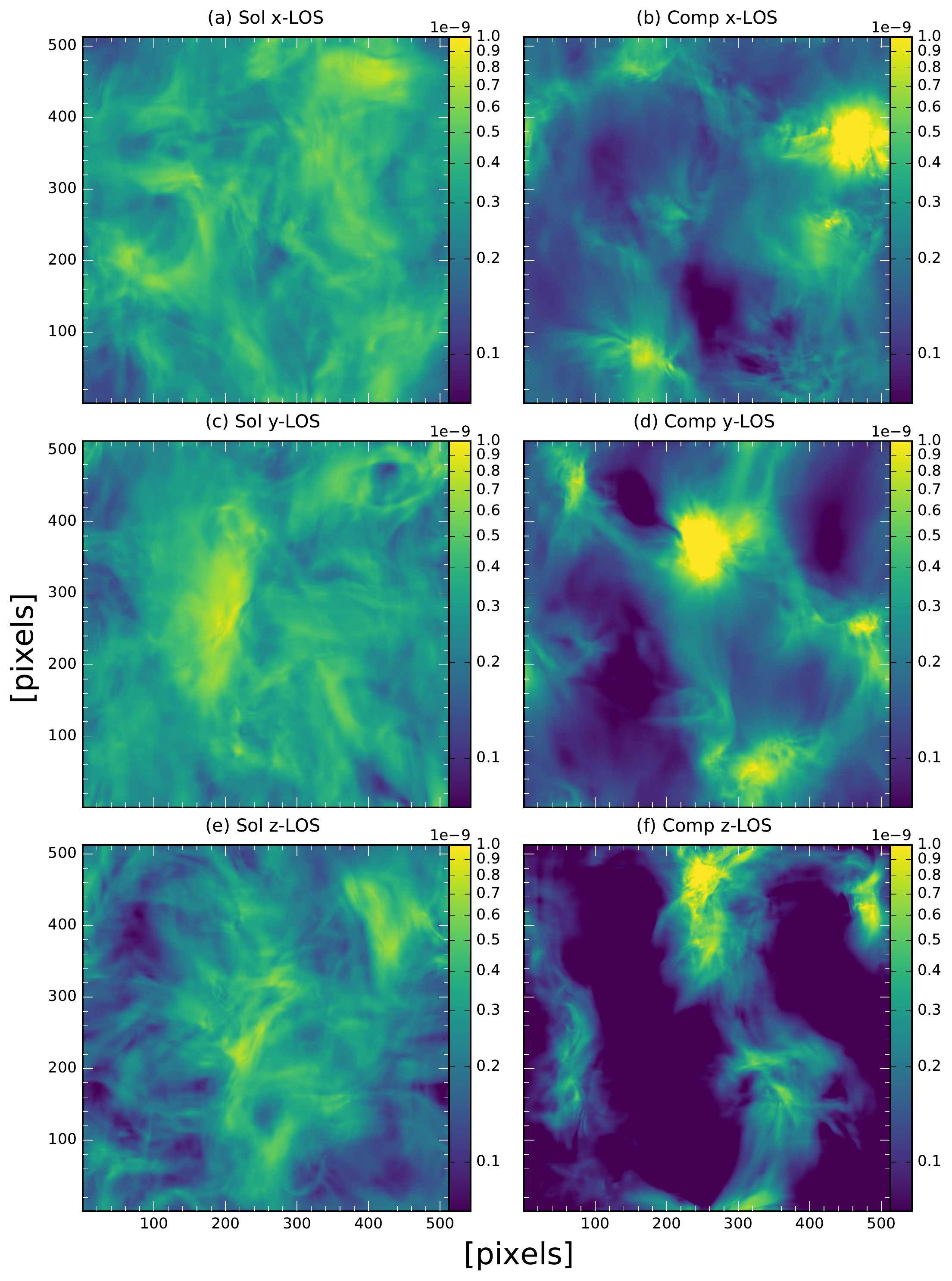}
\caption{Mock synchrotron intensity images for the simulations with solenoidal driving (left column) and compressive driving (right column) of MHD turbulence. The synchrotron image for a line of sight along the x-axis is shown in the top row, along the y-axis in the middle row, and along the z-axis in the bottom row. Note that a logarithmic colour scale is used for both simulations. The units specified are proportional to synchrotron intensity, see Equation \ref{mock_sync}.}
\label{fig:sync_image}
\end{center}
\end{figure*}

In Figure \ref{fig:sync_image} we present example images of synchrotron intensity obtained for the simulation with solenoidal driving (left, hereafter referred to as the solenoidal simulation), and the simulation with compressive driving (right, hereafter referred to as the compressive simulation), for each line of sight, and $t = 2t_{\mathrm{turb}}$. We find that the compressive simulation tends to have very bright, localised regions of synchrotron emission, but much smaller values of synchrotron intensity away from these few localised areas. On the other hand, bright synchrotron intensity regions appear to be less localised for the solenoidal simulation, and there is a smaller spread in values of synchrotron intensity compared to the compressive simulation. These differences are caused by the formation of very strong shocks in the compressive simulation, which provide strong amplification of the magnetic field in localised regions, whereas in the solenoidal simulation the magnetic field is primarily amplified by tangling and twisting of the magnetic field throughout the simulation cube, due to the greater vorticity in this simulation \citep{Federrath2011b}. 

To ensure that our mock synchrotron intensity images can be directly compared to observations, we normalised each synchrotron image by dividing by the mean synchrotron intensity of the image, and then calculated the logarithm of the normalised synchrotron intensity image. We call the log-normalised synchrotron intensity $\mathcal{I} = \log_{10}{(I/<I>)}$, where $<I>$ is the mean synchrotron intensity in the image. This quantity can be easily calculated for observed synchrotron intensity images, providing a means through which we can compare simulations and observations without needing to know the average magnetic field perpendicular to the line of sight, the cosmic ray electron density, or the depth along the line of sight of the volume from which synchrotron emission has been observed. 

To investigate whether it is possible to distinguish between different types of turbulent driving, we calculate the moments of $\mathcal{I}$, namely the mean $\mu_{\mathcal{I}}$, standard deviation $\sigma_{\mathcal{I}}$, biased skewness $\gamma_{\mathcal{I}}$, and biased Fisher kurtosis $\beta_{\mathcal{I}}$, given by:
\begin{align}
\mu_{\mathcal{I}} &= \frac{1}{N} \sum_{i=1}^{N} \mathcal{I}_i , \label{mean_i} \\
\sigma_{\mathcal{I}} &= \sqrt{\frac{1}{N} \sum_{i=1}^{N} (\mathcal{I}_i - \mu_{\mathcal{I}})^2} , \label{stdev_i} \\
\gamma_{\mathcal{I}} &= \frac{1}{N}  \sum_{i=1}^{N} \left(\frac{\mathcal{I}_i - \mu_{\mathcal{I}}} {\sigma_{\mathcal{I}}}\right)^3 , \text{and} \label{skew_i} \\
\beta_{\mathcal{I}} &=  \frac{1}{N}  \sum_{i=1}^{N} \left(\frac{\mathcal{I}_i - \mu_{\mathcal{I}}} {\sigma_{\mathcal{I}}}\right)^4 - 3. \label{kurt_i}
\end{align}
In these equations, $N$ is the total number of pixels in the image, and we sum over all pixels. Positive values of skewness indicate that the PDF of $\mathcal{I}$ has an extended tail towards large values of $\mathcal{I}$, and negative values of skewness indicate an extended tail towards small values of $\mathcal{I}$. Positive values of kurtosis indicate a PDF that is more peaked than a Gaussian, and negative values indicate a PDF that is flatter than a Gaussian. We study these moments as they have been applied to various observable quantities  \citep{Kowal2007, Burkhart2009, Burkhart2010, Federrath2010, Gaensler2011, Burkhart2012} due to their sensitivity to properties of turbulence.


\section{Results} \label{sec_results}
\begin{figure*}
\begin{center}
\includegraphics[scale=0.7]{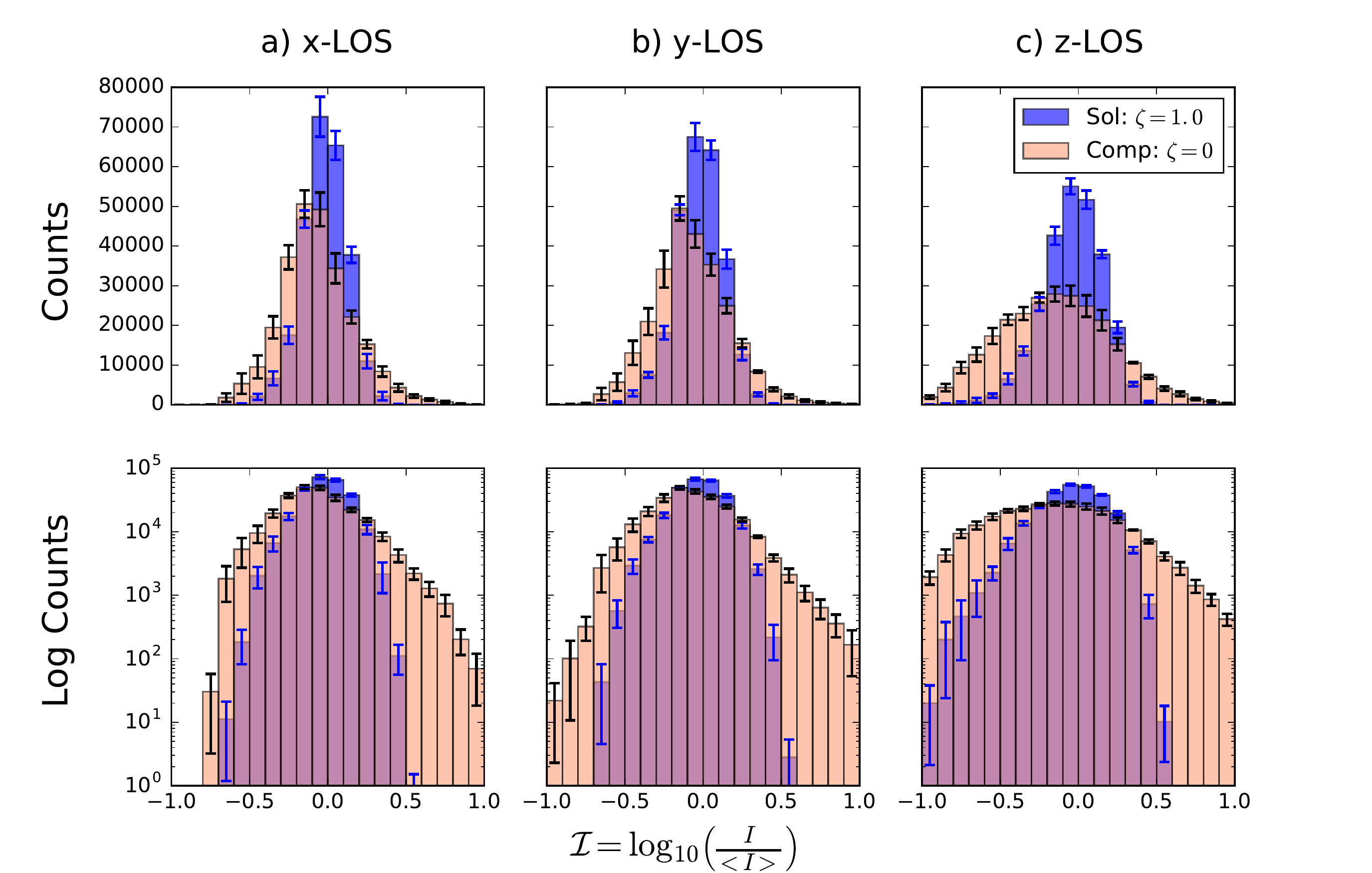}
\caption{Histograms of the time-averaged, log-normalised synchrotron intensity for lines of sight along a) the x-axis, b) the y-axis and c) the z-axis. The bottom row depicts the same PDFs as the top row, but with a logarithmic y-axis. The error bars denote the standard error of the mean number of counts in that bin, calculated over the different snapshots of the simulation. The simulation with solenoidal driving ($\zeta = 1$) is shown in blue, with blue error bars, and the simulation with compressive driving ($\zeta = 0$) is shown in orange, with black error bars (overlapping areas appear purple).}
\label{fig:sync_pdfs}
\end{center}
\end{figure*}

In Figure \ref{fig:sync_pdfs} we show the PDFs of $\mathcal{I}$ for the simulations with solenoidal ($\zeta = 1.0$, blue) and compressive ($\zeta = 0$, orange) driving, for lines of sight along the x, y, and z axes. The PDFs shown on the top row have a linear y-axis, whereas those on the bottom row have a logarithmic y-axis, to emphasise the tails of the distribution. These PDFs have been time-averaged, so that the number of counts in each bin is the average number of counts in that bin, calculated for the different realisations, for that line of sight. The uncertainties shown for each bin represent the standard error of the mean number of counts in the bin.

\begin{figure*}
\begin{center}
\includegraphics[scale=0.8]{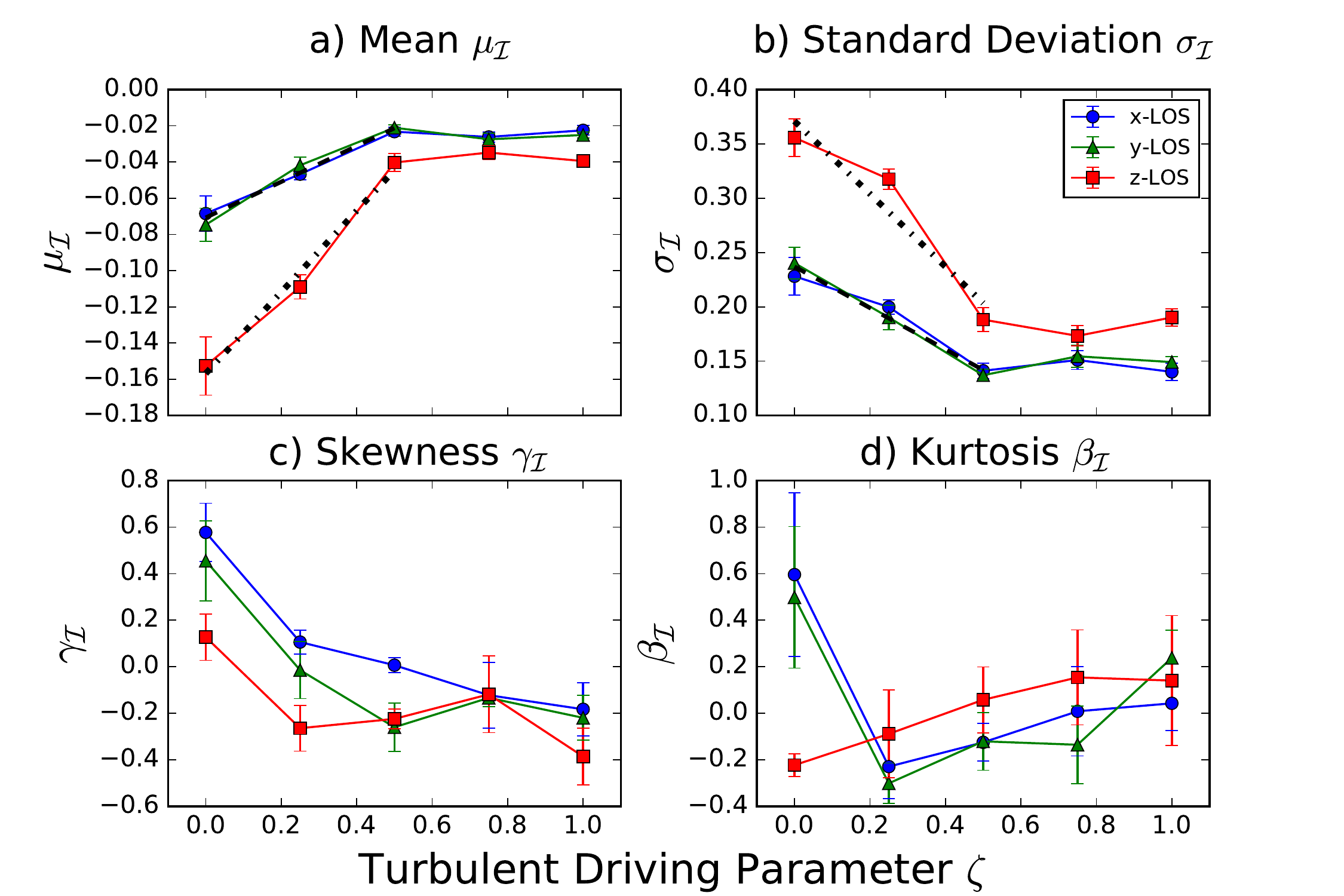}
\caption{The time-averaged a) mean, b) standard deviation, c) skewness, and d) kurtosis of the log-normalised synchrotron intensity maps as a function of the turbulent driving parameter $\zeta$. The error bars denote the standard error of the mean of the statistic for that simulation, calculated over the different snapshots of the simulation. The statistics calculated for lines of sight along the x-axis are shown as blue circles, along the y-axis as green triangles, and along the z-axis as red squares. The lines of best fit are drawn as dashed and dot-dashed lines, for lines of sight that are perpendicular or parallel to the mean magnetic field, respectively.}
\label{fig:sync_stats}
\end{center}
\end{figure*}

We find that the solenoidal simulation has a larger mean value of $\mathcal{I}$ for all lines of sight, and that the PDFs for the compressive simulation have a larger spread than the PDFs for the solenoidal simulation for all lines of sight. This suggests that it should be possible to distinguish between different values of $\zeta$, and different types of turbulent driving, by calculating the mean and standard deviation of $\mathcal{I}$. We also observe that the difference between the PDFs of the compressive and solenoidal simulations is most pronounced for a line of sight along the z axis, which is parallel to the mean magnetic field. To test whether these conclusions are robust, in Figure \ref{fig:sync_stats} we plot the time-averaged mean, standard deviation, skewness and kurtosis of $\mathcal{I}$ as a function of $\zeta$, for all lines of sight. The uncertainty bars shown represent the standard error of the mean.

We see that the mean value of $\mathcal{I}$ increases with increasing $\zeta$ up to $\zeta = 0.5$, and then plateaus, for every line of sight. For the standard deviation, we find that $\sigma_{\mathcal{I}}$ decreases with increasing $\zeta$, up to $\zeta = 0.5$, and then plateaus, again for every line of sight. We conclude that the mean and standard deviation of $\mathcal{I}$ are sensitive tracers of $\zeta$, for values of $\zeta$ between $0$ and $0.5$. We also find that the mean and standard deviation are most sensitive to changes in $\zeta$ if the line of sight is parallel to the mean magnetic field.

We calculated lines of best-fit for the mean and standard deviation of the log-normalised intensity, for the cases where the lines of sight are parallel and perpendicular to the mean magnetic field. These lines were produced for $0 \le \zeta \le 0.5$, and are shown as a black dashed line for lines of sight perpendicular to the mean magnetic field, and as a dot-dashed line for lines of sight parallel to the mean magnetic field. We assume that the mean and standard deviation plateau for $0.5 \le \zeta \le 1.0$. From these lines, we are able to express $\zeta$ as a function of the mean or standard deviation. For lines of sight perpendicular to the mean magnetic field:
\begin{align}
\zeta(\mu_{\mathcal{I}}) &= 4.4\mu_{\mathcal{I}} + 0.7\\
\zeta(\sigma_{\mathcal{I}}) &= -3.0\sigma_{\mathcal{I}} + 1.1
\end{align}
For lines of sight parallel to the mean magnetic field:
\begin{align}
\zeta(\mu_{\mathcal{I}}) &= 10.1\mu_{\mathcal{I}} + 0.7\\
\zeta(\sigma_{\mathcal{I}}) &= -5.3\sigma_{\mathcal{I}} + 1.2
\end{align}
In Appendix \ref{mach_dep} we investigate how the mean and standard deviation of the log-normalised intensity depend on the sonic and Alfv\'enic Mach numbers, to determine how robust our lines of best-fit are to changes in the Mach numbers. We find that the mean and standard deviation have no dependence on the sonic Mach number, and depend on the Alfv\'enic Mach number in a monotonic fashion. Assuming that the Alfv\'enic Mach number of an observed molecular cloud is similar to the Alfv\'enic Mach numbers in our simulations, these formulae can hence be used to determine $\zeta \in [0,0.5]$ from the measured log-normalised synchrotron intensity, for any sonic Mach number.

The bottom panels of Figure \ref{fig:sync_stats} show that the skewness and kurtosis of $\mathcal{I}$ do not depend as strongly on $\zeta$ as the mean and standard deviation of $\mathcal{I}$. Whilst the skewness may decrease with $\zeta$ for values of $\zeta$ between $0$ and $0.25$, there does not appear to be any significant dependence of the skewness on $\zeta$ for values of $\zeta$ above $0.25$. There also does not appear to be any significant dependence of the kurtosis on $\zeta$. This may be because the skewness and kurtosis are high-order moments, and can be sensitive to random fluctuations in the simulations, i.e., low-number statistics, as shown by the large error bars on these statistics. It is possible that with larger simulations, the skewness and kurtosis could be determined to greater precision, and a dependence on $\zeta$ revealed. 

Finally, we note that we have calculated the structure function slope and integrated quadrupole ratio modulus statistics, as calculated by \cite{Herron2016}, and found that these statistics had no correlation with $\zeta$.


\section{Discussion} \label{sec_discuss}
\subsection{Effects of Finite Noise and Angular Resolution} \label{sec_discuss_obs}
To investigate whether current telescopes will be able to apply our method to determine the turbulent driving parameter $\zeta$, we consider how the mean and standard deviation of the log-normalised synchrotron intensity are affected by the observational noise and angular resolution. Our method is the same as that described by \cite{Herron2016}, which we briefly summarise here. We add randomly generated Gaussian noise to the synchrotron intensity maps, where the standard deviation of the Gaussian distribution is a fraction of the median synchrotron intensity. This fraction was varied to control the strength of the added noise. To simulate the effect of angular resolution on these statistics, the synthetic synchrotron intensity image was convolved with a Gaussian beam, and the size of this Gaussian was varied to control the smoothing scale. We simulate the most realistic case where there is both noise and finite angular resolution by first adding noise to the simulated synchrotron map, and then smoothing to the desired resolution.

\begin{figure*}
\begin{center}
\includegraphics[scale=0.8]{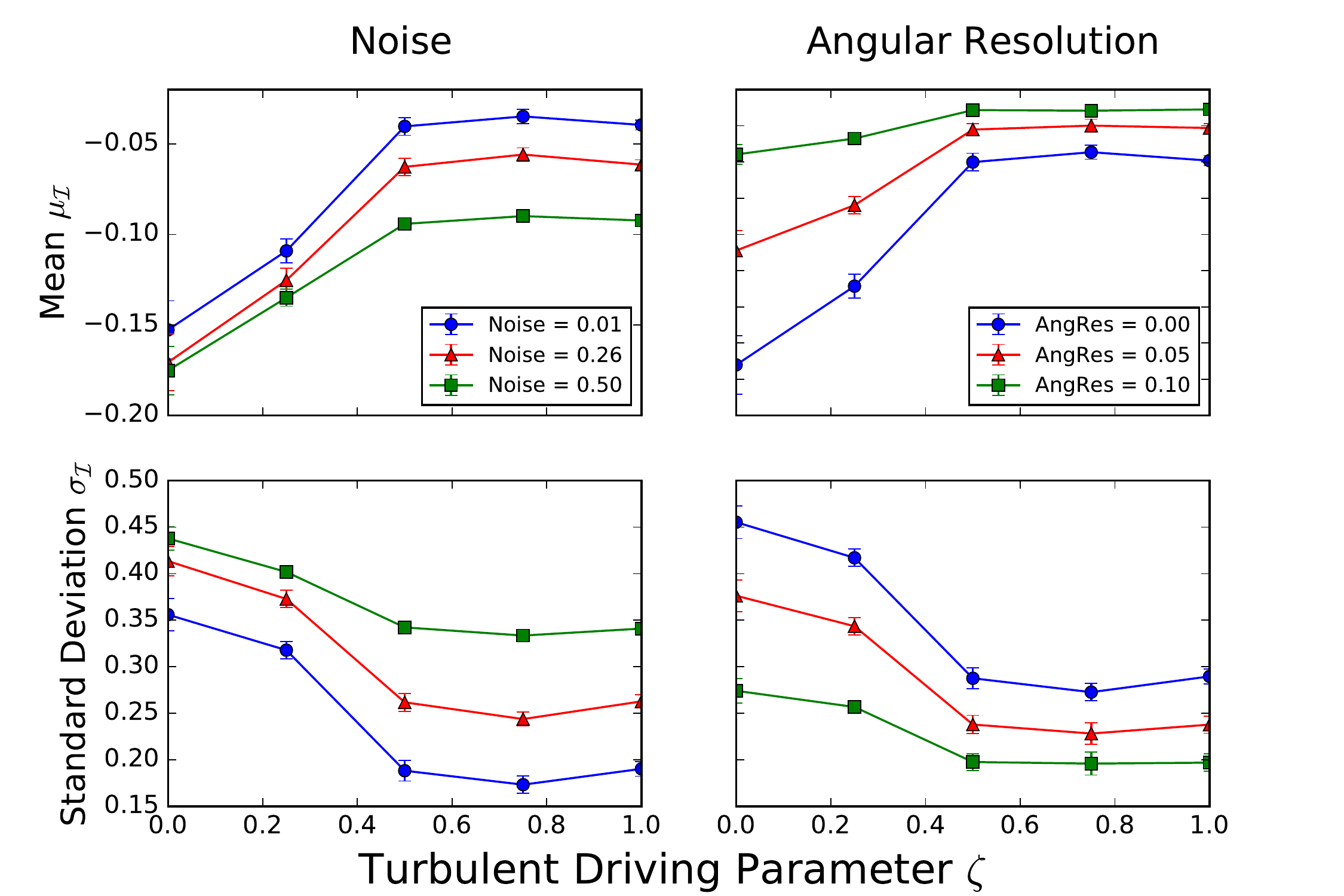}
\caption{The time-averaged mean (top row) and standard deviation (bottom row) of the log-normalised synchrotron intensity for a line of sight parallel to the mean magnetic field, as a function of the turbulent driving parameter $\zeta$, for varying noise levels (left column) and angular resolutions (right column). The noise levels specified are noise-to-signal ratios, and the angular resolutions specified express the smoothing scale as a fraction of the total size of a molecular cloud. The error bars denote the standard error of the mean of the statistic for that simulation, calculated over the different snapshots of the simulation.}
\label{fig:noise_dep}
\end{center}
\end{figure*}

In Figure \ref{fig:noise_dep} we plot the time-averaged mean and standard deviation of the log-normalised synchrotron intensity against the turbulent driving parameter $\zeta$, for a line of sight parallel to the mean magnetic field. In the left column we show how the relationship between these statistics and $\zeta$ varies as the noise level is increased, and in the right column we show how the relationship depends on the smoothing scale. The noise levels specified are noise-to-signal ratios, and the angular resolutions specified express the standard deviation of the Gaussian used to smooth the synchrotron intensity map as a fraction of the width of the image. This expresses the angular resolution as a fraction of the total size of the molecular cloud.

We find that for all noise levels and angular resolutions studied, there is always a correlation between the mean and standard deviation of the log-normalised synchrotron intensity and the turbulent driving parameter, for $\zeta \in [0,0.5]$. For larger noise levels, or poorer angular resolution, the mean and standard deviation become less sensitive to $\zeta$, but this is not significant enough to erase all dependence of these statistics on the turbulent driving. 

\begin{figure*}
\begin{center}
\includegraphics[scale=0.8]{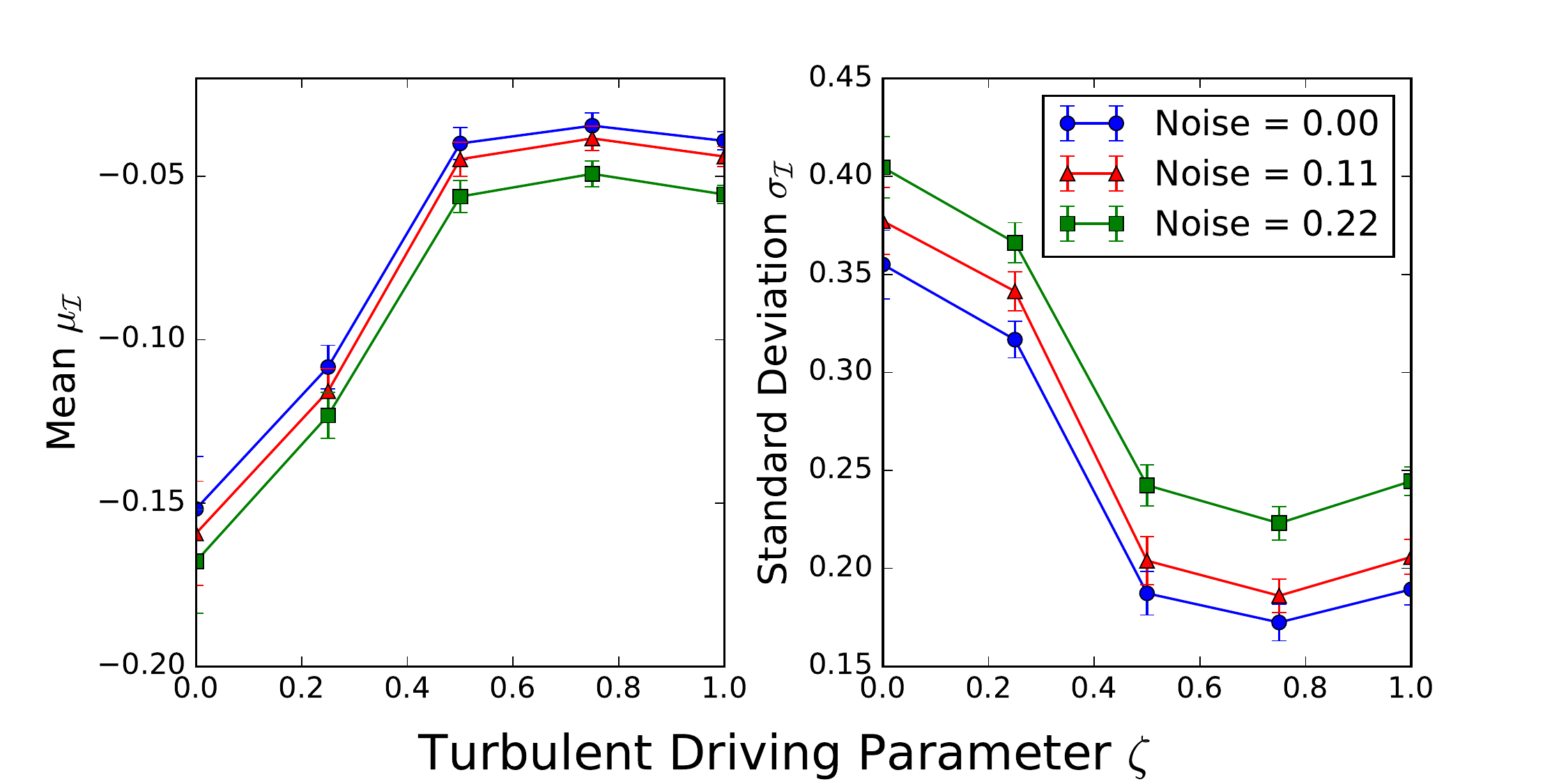}
\caption{The time-averaged mean (left) and standard deviation (right) of the log-normalised synchrotron intensity for a line of sight parallel to the mean magnetic field, as a function of the turbulent driving parameter $\zeta$, for varying noise levels, at a fixed angular resolution of $3$ pixels FWHM. The noise levels specified are noise-to-signal ratios. The error bars denote the standard error of the mean of the statistic for that simulation, calculated over the different snapshots of the simulation.}
\label{fig:noise_ang_dep}
\end{center}
\end{figure*}

In Figure \ref{fig:noise_ang_dep} we plot the time-averaged mean and standard deviation of the log-normalised synchrotron intensity against the turbulent driving parameter $\zeta$ for the case where there is both noise and finite angular resolution. The synchrotron maps were smoothed to have a FWHM of $3$ pixels, and the noise levels are specified as the noise-to-signal ratios in the image after smoothing. We find that the relationship between the mean or standard deviation and the turbulent driving parameter persists even in the presence of noise and finite angular  resolution. Signal-to-noise ratios above $5$, and angular resolutions that are smaller than $5\%$ of the total cloud diameter, should be sufficient to permit $\zeta$ to be determined from measured statistics. 

Finally, we note that changes in the noise level or angular resolution introduce quantitative changes in the relations between $\zeta$ and $\mu_{\mathcal{I}}$, as well as between $\zeta$ and $\sigma_{\mathcal{I}}$, and so these observational effects must be taken into account before attempting to use measured statistics to determine $\zeta$. 


\subsection{Observability of Synchrotron Emission from Molecular Clouds} \label{observability}
As discussed in Section \ref{Intro}, \cite{Protheroe2008} and \cite{Jones2011} expected to see synchrotron emission from the Sagittarius B molecular cloud, but did not find any detections in their observations. This implies that the degree to which cosmic rays penetrate this molecular cloud is lower than expected. 

\cite{Gabici2007} and \cite{Gabici2009} investigate the propagation of cosmic ray protons and electrons within molecular clouds of different sizes, masses, and magnetic field strengths. In particular, \cite{Gabici2009} found that cosmic ray electrons with energies between $100$ MeV and a few hundred TeV are capable of travelling through giant molecular clouds, because the propagation time for these electrons to travel through the cloud is shorter than the timescale for energy loss. This is supported by \cite{Boettcher2013}, who found that cosmic rays can pass through molecular clouds several times before losing a significant amount of energy. Cosmic ray electrons in the energy range of $100$ MeV to $100$ TeV emit synchrotron emission at frequencies between $0.5$ MHz and x-ray frequencies, and so for radio observations at MHz or GHz frequencies, the density of cosmic rays emitting the observed synchrotron radiation should be the same inside and outside a giant molecular cloud. However, the Sagittarius B molecular cloud is smaller, more massive, and has a stronger magnetic field than typical giant molecular clouds, and so for this cloud the energy loss timescale is smaller than the propagation timescale for all cosmic ray electrons \citep{Gabici2009}. This means that cosmic ray electrons are completely excluded from the Sagittarius B molecular cloud, and synchrotron emission is only emitted by secondary electrons that are produced in the molecular cloud by collisions between very high energy cosmic ray protons and the molecular gas \citep{Brown1977, Marscher1978}. As discussed by \cite{Protheroe2008} and \cite{Jones2011}, their non-detection of synchrotron emission from secondary electrons may indicate that it is more difficult for cosmic ray protons to penetrate Sagittarius B than expected.

The main factor hindering the detection of synchrotron emission from the giant molecular clouds we consider is confusion between the faint synchrotron emission originating from within molecular clouds and emission from other parts of the ISM (i.e., foreground and background emission arising from outside the molecular cloud needs to be separated from the signal we are interested in). While it may be possible to separate synchrotron emission from within a molecular cloud by comparing the statistics and spectra of synchrotron emission coincident and outside a molecular cloud, the low surface-brightness of this emission is the most immediate challenge to overcome, as otherwise our method cannot be applied to future observations.

To estimate whether current or upcoming telescopes will be able to detect diffuse synchrotron emission from giant molecular clouds, we consider the Perseus molecular cloud, which is $200$--$350$ pc away \citep{Herbig1983}, and has an angular size of $6^{\circ} \times 3^{\circ}$ on the sky. We assume a spectral index $\alpha = -0.7$, and that the depth of the cloud is approximately $20$ pc (similar to the extent of the Perseus cloud across the sky). We also assume a magnetic field strength of $10 \, \mu$G throughout the cloud, based on the observations of \cite{Crutcher1993}, who measured the component of the magnetic field parallel to the line of sight to be $19 \, \mu$G in Barnard 1, a dense molecular core within the Perseus cloud. As the magnetic field strength is larger in dense cores than in the more diffuse regions of giant molecular clouds, we estimate that the magnetic field strength will be approximately $10 \, \mu$G in the diffuse regions of the Perseus cloud. This assumption is consistent with the findings of \cite{CrutcherEtAl2010}, who found that the maximum magnetic field strength in clouds with densities below $300 \, \text{cm}^{-3}$ is $10 \, \mu$G. Under these assumptions, we estimate a synchrotron brightness of $1.7$ mK at $1.4$ GHz, at an angular resolution of $9'$. We also assumed that primary cosmic ray electrons penetrate the cloud, and that the cosmic ray density is similar to that at Earth. For comparison, we estimate the synchrotron intensity due to emission between the Perseus cloud and an observer to be $3.4$ mK, assuming an interstellar magnetic field strength perpendicular to the line of sight of $3 \mu$G \citep{Sun2008}. The synchrotron emission from behind the Perseus cloud may be much brighter than the foreground emission. For instance, \cite{Tibbs2013} measured an RMS confusion level of $20$ mJy beam$^{-1}$ at 1.4 GHz, and an angular resolution of $9'$, in the direction of the Perseus cloud, corresponding to a brightness temperature of $43$ mK. 

The Green Bank Telescope (GBT) should be capable of reaching a thermal sensitivity below $1.7$ mK at this angular resolution, and so may be able to observe synchrotron emission from the Perseus molecular cloud. This assumes that the synchrotron emission from the molecular cloud can be reliably separated from the foreground and background emission. To be able to apply our method of determining $\zeta$ to the Perseus molecular cloud, this corresponds to a required sensitivity of $0.3$ mK at an angular resolution of approximately $9'$. The confusion level is significantly larger than this and so it may not be possible to apply our method to observations of the Perseus molecular cloud performed with just the GBT. However, if the same region is observed with the Very Large Array (VLA) in D and C configuration, then the high-resolution interferometer data can be used to detect the point sources that contribute to the confusion level in the GBT observations. By modelling the appearance of these sources at the GBT resolution, it is possible to subtract these sources from the GBT image, lowering the confusion level to approximately $0.1$ mK. It should then be possible to detect synchrotron emission from the Perseus cloud at sufficient signal-to-noise and angular resolution to be able to estimate $\zeta$.

\subsection{Limitations and Extensions of Our Method} \label{applicability}
We have found that it is possible to use the mean and standard deviation of the log-normalised synchrotron intensity, $\mathcal{I}$, to constrain the value of $\zeta$, provided that the value of $\zeta$ is between $0$ and $0.5$. This essentially allows us to constrain how compressive the driving of an observed turbulent region is, and whether star formation or supernova explosions have had a significant impact on the dynamics of the gas. However, we are unable to distinguish between values of $\zeta$ that lie between $0.5$ and $1$. This is expected, as for $\zeta$ values above $0.5$ at least two thirds of the energy that is injected into the turbulence is in solenoidal motions, which should dominate the statistics of the turbulence \citep{Federrath2010}. Mixed driving ($\zeta=0.5$) and purely solenoidal driving ($\zeta=1$) are very similar with respect to the amount of compressibility induced by the driving, so we do not expect a significant variation of the synchrotron statistics for $\zeta=[0.5,1]$. 

We also find that the relationship between the mean and standard deviation of $\mathcal{I}$, and $\zeta$, depends on the orientation of the line of sight relative to the mean magnetic field (see Section \ref{sec_results}). In particular, the mean and standard deviation are most sensitive to $\zeta$ for a line of sight that is parallel to the mean magnetic field. This means that the relative angle between the line of sight and the mean magnetic field should be determined prior to measuring $\zeta$, for example through observations of Zeeman splitting (to determine the component of the magnetic field parallel to the line of sight), and polarised thermal dust emission or synchrotron emission (to determine the component of the magnetic field perpendicular to the line of sight). Alternatively, this relative angle could be thought of as a free parameter, and determined using statistics of synchrotron emission whilst $\zeta$ is being determined. This would involve calculating synchrotron statistics for different combinations of the relative angle and $\zeta$, to find the particular combination that produces values of the statistics that are closest to those observed (similar to using a Markov Chain Monte Carlo program to determine the best-fit parameter values for a model).

As the synchrotron intensity that we observe is brightest when the magnetic field is perpendicular to the line of sight, we may not be able to use the mean and standard deviation of $\mathcal{I}$ to constrain $\zeta$ along lines of sight for which the mean and standard deviation attain maximum sensitivity to $\zeta$. This should not pose a problem for measurements of $\zeta$, as even if the line of sight is perpendicular to the mean magnetic field, the mean and standard deviation of the log-normalised intensity are significantly correlated with $\zeta$, and so should be sufficient to constrain $\zeta$. We believe that this dependence of the mean and standard deviation of $\mathcal{I}$ on the line of sight arises because the magnetic field acts to remove the signature of compressive turbulent driving, possibly by constraining plasma into filaments, and redirecting plasma motions. This is in agreement with the conclusions of \cite{Vestuto2003}, who found that increased magnetic field strengths in their simulations led to stronger solenoidal motions. There is an interesting comparison here to the findings of \cite{Herron2016}, who showed that their favoured statistics of synchrotron intensity, namely the structure function slope and integrated quadrupole ratio modulus, were more sensitive to the Alfv\'enic Mach number for lines of sight that were perpendicular to the mean magnetic field. This is because the structure function slope and integrated quadrupole ratio modulus probe observational signatures of the magnetic field, and hence are more sensitive to the properties of the turbulence when the line of sight is perpendicular to the mean magnetic field, whereas the observational signatures we probe are erased by the magnetic field.

As shown by \cite{Herron2016}, properties of turbulence such as the sonic and Alfv\'enic Mach numbers can influence statistics of synchrotron intensity, and so it is possible that the mean and standard deviation of $\mathcal{I}$ depend not only on $\zeta$ and the orientation of the line of sight relative to the mean magnetic field, but also on these Mach numbers. In Appendix \ref{mach_dep} we investigate this for the simulations and synchrotron maps produced by \cite{Herron2016}, and find that the mean and standard deviation of the log-normalised synchrotron intensity do not depend on the sonic Mach number, and depend monotonically on the Alfv\'enic Mach number, such that the amplitude of these statistics decreases as the Alfv\'enic Mach number decreases. This supports our belief that increased magnetic field strength (lower Alfv\'enic Mach number) removes the signature of compressive turbulent driving. 

The consequences of these findings is that it should be possible to determine the turbulent driving parameter from these statistics of synchrotron intensity for turbulent media (not necessarily molecular clouds) of any sonic Mach number, provided the Alfv\'enic Mach number is approximately larger than one. For small Alfv\'enic Mach numbers, the mean and standard deviation should be less sensitive to $\zeta$ than at high Alfv\'enic Mach numbers, although future studies investigating the relationship between these statistics and $\zeta$ for sub-Alfv\'enic simulations will be required to confirm this. As such, the best-fit formulae presented in this paper should only be applied to turbulent regions with similar Alfv\'enic Mach numbers to our simulations, namely molecular clouds. However, if the relationship between our statistics and $\zeta$ is confirmed for lower Alfv\'enic Mach numbers, then our method can be applied to other phases of the interstellar medium, such as the warm ionised medium. Future studies will also be required to determine how the statistics change for Alfv\'enic Mach numbers greater than two. 


\section{Conclusions}

We have calculated the statistical moments of mock log-normalised synchrotron intensity images for magnetohydrodynamic simulations of molecular clouds, and compared these statistics to the turbulent driving parameter $\zeta$. We found that the mean and standard deviation of the log-normalised synchrotron intensity are sensitive to $\zeta$ for values of $\zeta$ between $0$ and $0.5$. These statistics can be used to constrain how compressive the driving of an observed emitting region is, provided the Alfv\'enic Mach number of the observed molecular cloud is similar to that of our simulations, and the angle of the mean magnetic field relative to the line of sight can be determined. Although the mean and standard deviation are most sensitive to $\zeta$ when the line of sight is parallel to the mean magnetic field, these statistics remain sufficiently sensitive to $\zeta$ for lines of sight perpendicular to the mean magnetic field, for which synchrotron emission is brightest. The mean and standard deviation are also sensitive to $\zeta$ for signal-to-noise ratios and angular resolutions attainable with current telescopes, and we suggest that it may be possible to observe diffuse synchrotron emission from the Perseus molecular cloud if observations with the Green Bank Telescope are combined with observations with the Very Large Array, permitting the application of this technique. Future studies are required to confirm whether the mean and standard deviation maintain their sensitivity to $\zeta$ for various Alfv\'enic Mach numbers, which would mean the methods described in this paper can be applied to other phases of the interstellar medium.

\section*{Acknowledgements}

C.~A.~H. thanks Ron Ekers and Mordecai-Mark Mac Low for useful discussions on the observability of synchrotron emission from molecular clouds. C.~A.~H. acknowledges financial support received via an Australian Postgraduate Award, and a Vice Chancellor's Research Scholarship awarded by the University of Sydney. C.~F. acknowledges funding provided by the Australian Research Council's Discovery Projects (grant ~DP150104329). B.~M.~G. acknowledges the support of the Natural Sciences and Engineering Research Council of Canada (NSERC) through grant RGPIN-2015-05948. N.~M.~M.-G. acknowledges the support of the Australian Research Council through grant FT150100024. B.~B. is supported by the NASA Einstein Postdoctoral Fellowship. The Dunlap Institute for Astronomy and Astrophysics is funded through an endowment established by the David Dunlap family and the University of Toronto. We gratefully acknowledge the J\"ulich Supercomputing Centre (grant hhd20), the Leibniz Rechenzentrum and the Gauss Centre for Supercomputing (grants~pr32lo, pr48pi and GCS Large-scale project~10391), the Partnership for Advanced Computing in Europe (PRACE grant pr89mu), the Australian National Computational Infrastructure (grant~ek9), and the Pawsey Supercomputing Centre with funding from the Australian Government and the Government of Western Australia.
The software used in this work was in part developed by the DOE-supported Flash Center for Computational Science at the University of Chicago. This research made use of APLpy, an open-source plotting package for Python hosted at http://aplpy.github.com.

\appendix
\section{Mach Number Dependence of Synchrotron Statistics} \label{mach_dep}
\begin{figure*}
\begin{center}
\includegraphics[scale=0.8]{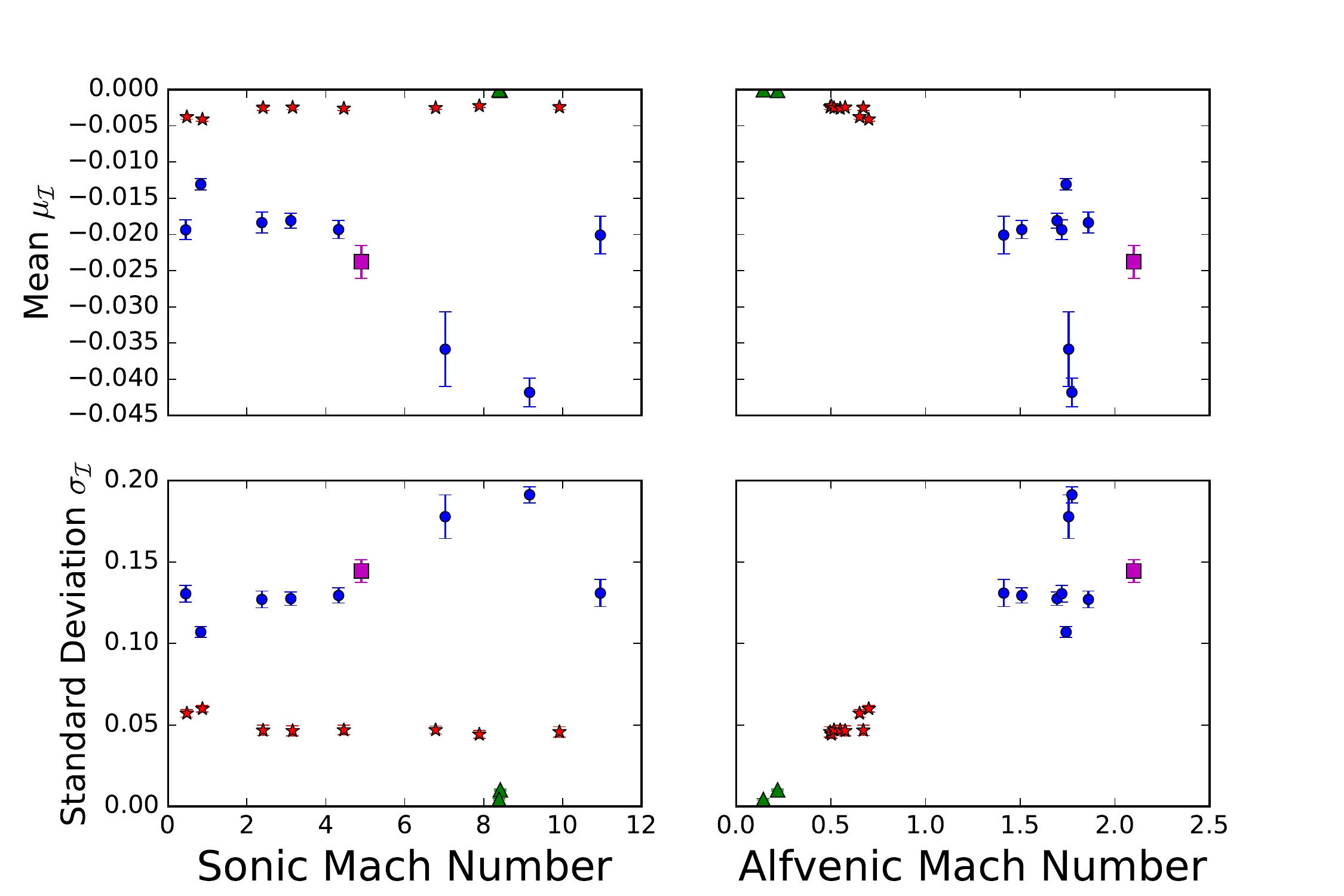}
\vspace{-1.0em}
\caption{The mean (top row) and standard deviation (bottom row) of the log-normalised synchrotron intensity maps as a function of the sonic Mach number (left column) and Alfv\'enic Mach number (right column), for the simulations described in \protect\cite{Herron2016}. Simulations with an initial magnetic field strength $B=0.1$ are shown in blue, those with $B=1$ are shown as red stars, and those with $B=3$ and $B=5$ are shown as green triangles. For these data points, the errors are calculated as described in \protect\cite{Herron2016}. The time-averaged statistics for our solenoidal simulation are shown for comparison, as a purple square, where the error bars shown denote the standard error of the mean.}
\label{fig:mach_dep}
\end{center}
\end{figure*}

It is important to determine whether the relationships between the mean and standard deviation of the log-normalised intensity, and the turbulent driving parameter $\zeta$ are robust to changes in the sonic and Alfv\'enic Mach numbers, so that the applicability of these relationships to arbitrary molecular clouds can be elucidated. To perform this study, we calculated the mean and standard deviation of the log-normalised synchrotron intensity for the solenoidally-driven simulations and synthetic synchrotron intensity observations described by \cite{Cho2003} and \cite{Herron2016}. These statistics were calculated for lines of sight perpendicular to the mean magnetic field of the simulation. The average value of these statistics for the two lines of sight perpendicular to the mean magnetic field are plotted against the sonic and Alfv\'enic Mach numbers of the simulations in Figure \ref{fig:mach_dep}. The error bars for these data points are calculated using the same method as in \cite{Herron2016}. The blue data points represent simulations with an initial magnetic field strength of $B=0.1$ (in simulation units), simulations with an initial magnetic field strength of $B=1$ are shown as red stars, and the simulations with initial magnetic field strengths of $B=3$ and $B=5$ are shown as green triangles. For comparison, we plot the mean and standard deviation of the log-normalised synchrotron intensity for our solenoidal simulation, averaged over lines of sight perpendicular to the mean magnetic field and the five temporal realisations of the simulation, as a purple square. The error bars represent the standard error of the mean calculated over the different realisations of the simulation, and the two lines of sight, and indicate that the statistics for our simulation are consistent with those of \cite{Herron2016}.

We find that the mean and standard deviation of the log-normalised synchrotron intensity do not have any dependence on the sonic Mach number for simulations with an Alfv\'enic Mach number $M_A \approx 0.6$. These statistics might have some dependence on the sonic Mach number for simulations with an Alfv\'enic Mach number $M_A \approx 1.7$, however given that two of these simulations have significantly different values of these statistics compared to the simulations with similar Alfv\'enic Mach numbers, we believe that these simulations may be outliers. If these two simulations are ignored, then the mean and standard deviation of the log-normalised intensity do not depend on sonic Mach number for simulations with $M_A \approx 1.7$ either. We also find that the mean and standard deviation of the log-normalised intensity depend on the Alfv\'enic Mach number monotonically, such that the absolute value of these statistics decreases as the Alfv\'enic Mach number decreases. This suggests that these statistics will lose their sensitivity to $\zeta$ as the magnetic field strength in the emitting region increases. Figure \ref{fig:mach_dep} also demonstrates that the mean and standard deviation of the log-normalised synchrotron intensity provide constraints on the Alfv\'enic Mach number, in addition to the statistics discussed by \cite{Herron2016}.



\bibliographystyle{/Users/chrisherron/Documents/PhD/My_Papers/mnras}
\bibliography{ref_list} 

\bsp	
\label{lastpage}
\end{document}